## Journal of Geophysical Research: Space Physics

# An altitude and distance correction to the source fluence distribution of TGFs

R. S. Nisi[1,2], N. Østgaard[1,2], T. Gjesteland[1,2], and A. B. Collier[3]

[1]Department of Physics and Technology, University of Bergen, Bergen, Norway, [2]Birkeland Center for Space Science, Bergen, Norway, [3]School of Chemistry and Physics, University of KwaZulu-Natal, Durban, South Africa

**Abstract** The source fluence distribution of terrestrial gamma ray flashes (TGFs) has been extensively discussed in recent years, but few have considered how the TGF fluence distribution at the source, as estimated from satellite measurements, depends on the distance from satellite foot point and assumed production altitude. As the absorption of the TGF photons increases significantly with lower source altitude and larger distance between the source and the observing satellite, these might be important factors. We have addressed the issue by using the tropopause pressure distribution as an approximation of the TGF production altitude distribution and World Wide Lightning Location Network spheric measurements to determine the distance. The study is made possible by the increased number of Ramaty High Energy Solar Spectroscopic Imager (RHESSI) TGFs found in the second catalog of the RHESSI data. One find is that the TGF/lightning ratio for the tropics probably has an annual variability due to an annual variability in the Dobson-Brewer circulation. The main result is an indication that the altitude distribution and distance should be considered when investigating the source fluence distribution of TGFs, as this leads to a softening of the inferred distribution of source brightness.

## 1. Introduction

Ever since the discovery of terrestrial gamma ray flashes(TGFs) about 20 years ago [*Fishman et al.*, 1994], they have been known to be connected to thunderstorms. However, the predicted altitude of TGF production has changed significantly over the years, from more than 30 km [*Fishman et al.*, 1994] to less than 15 km [*Dwyer et al.*, 2012; *Dwyer and Smith*, 2005; *Carlson et al.*, 2007]. TGFs are now expected to be produced within the thunderstorm. As the tropopause constitutes the upper limit for most clouds and the positive layer of a thundercloud can be expected to be close to the top of the thundercloud, the tropopause pressure distribution has been used as an approximation for the TGF production altitude distribution. Some clouds with exceptionally strong updraft might overshoot the tropopause. *Liu and Zipser* [2005] found that around 9000 deep convective systems were overshooting the tropopause, as defined in this paper, during 5 years. In the same years, Lightning Imaging Sensor (LIS) onboard the same satellite measured around 1.2 million thunderstorms. This means that around 0.8% of all thunderstorms within ± 35° latitude has overshooting clouds. Even if the overshooting thunderstorms might have 2–4 times as many flashes as more shallow thunderstorms [*Liu and Zipser*, 2005], this indicate that only a small population of events are from overshooting clouds.

The pressure and altitude of the tropopause are highly variable in both space and time. They varies with surface temperature, solar radiation, updrafts, and atmospheric water content. The general global pattern is a high tropopause in the tropics (up to about 16 km altitude) that becomes lower with increasing latitude (around 8 km altitude at the poles). For a given location, the annual mean usually differs up to 0.5 km between years, while the monthly mean varies around 1 km over a year [*Seidel and Randel*, 2006]. The day-to-day variability for a given location has a mean of 45 hPa or 1.4 km but can be as much as 2 km [*Seidel and Randel*, 2006]. The altitude and pressure of the tropopause are important for establishing how much air the gamma rays have to go through to reach satellite altitude. Furthermore, this can help us to estimate more precisely how many photons are produced in each TGF.

The first measurements of TGFs were made by the Burst and Transient Source Experiment on board the Compton Gamma Ray Observatory [*Fishman et al.*, 1994]. Later observations were made by the Reuven Ramaty High Energy Solar Spectroscopic Imager (RHESSI), the Gamma-ray Burst Monitor on board the Fermi

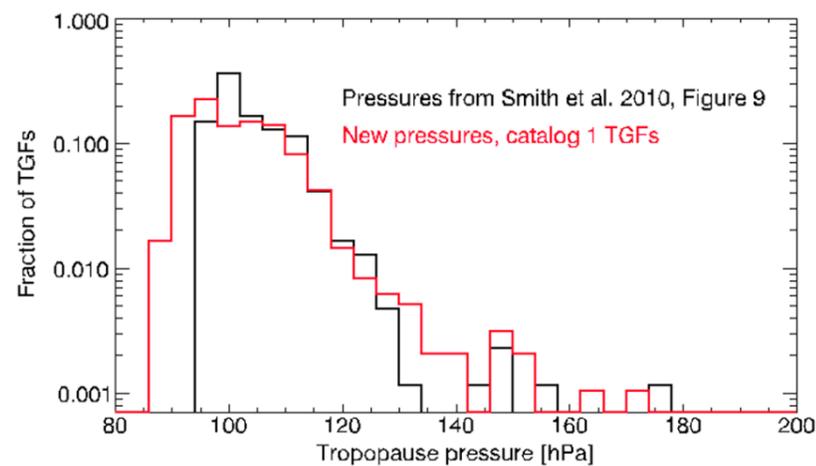

**Figure 1.** Tropopause pressure at the time and place of TGFs. *Smith et al.* [2010] used monthly tropopause values and the closest point in space, whereas we use the mean pressure within 400 km around the satellite foot point (closest point to the geolocated lightning where we have WWLLN match) and the closest time up to 3 h before or after the time of the TGFs. The figure demonstrates that our method gives a wider pressure distribution than *Smith et al.* [2010].

Gamma-ray Space Telescope and the Astrorivelatore Gamma a Immagini Leggero satellite [*Dwyer et al.*, 2012].

The first catalog of TGFs measured by RHESSI was reported by *Grefenstette et al.* [2009], and later, a new search algorithm for finding TGFs in the data set was developed by *Gjesteland et al.* [2012] (second catalog). For data between 2002 and 2011 the new search algorithm found around 2500 TGFs compared to the 900 in the first catalog. This increase in numbers also give us a larger number of TGFs with a corresponding spheric match from the World Wide Lightning Location Network (WWLLN) as described in *Gjesteland et al.* [2012], especially after 2007 when WWLLN increased their detection efficiency [*Collier et al.*, 2011].

In this paper we first describe the method used to find the tropopause at the time and place of the TGFs and compare the method to the previous work by *Smith et al.* [2010]. We will then compare the first and second catalog of RHESSI TGFs. Finally, we will use the tropopause distribution as an approximation of the TGF production altitude distribution and, when adjusting for the altitude and distance, estimate the fluence distribution at the source. *Carlson et al.* [2012] addressed how the source fluence distribution looks like but only addressed the source distance from the satellite observation using a constant production altitude. This paper expand that work to also include the estimated relative altitude distribution of the TGF source.

## 2. Method

We used tropopause pressure data from the National Centers for Environmental Prediction (NCEP)/National Center for Atmospheric Research (NCAR) 40 years reanalysis from NOAA/Earth System Research Laboratory (ESRL), an empirical model based on weather data *Kalnay et al.* [1996]. The NCEP/NCAR Reanalysis data are provided by the NOAA/OAR/ESRL Physical Science Division, Boulder, Colorado, USA, from their Web site at http://www.esrl.noaa.gov/psd/. The Reanalysis data provided are on a horizontal grid of 2.5 × 2.5°, and at several pressure levels including the tropopause level. The tropopause is defined by the standard definition from the World Meteorological Organization's Commission of Aerology [*Seidel and Randel*, 2006], and the model provides the temperature and pressure at the tropopause. We choose to use the pressure as a measure of the tropopause altitude, as this enables us to directly find the amount of air above the tropopause. The NCEP/NCAR reanalysis has been found to give slightly too high tropopause pressures in the tropics. In an analysis by *Randel et al.* [2000] they found a mean error of about 4 hPa with little or no seasonal dependence for 26 stations, while *Kaladis et al.* [2001] found a small annual variability using only one station. The error is assumed to be related to how the satellite measurements are implemented into the model [*Kaladis et al.*, 2001].

In all our analysis we use the tropopause pressure as provided by NCEP/NCAR. To improve the readability for readers not familiar with the use of pressure coordinates, we sometimes refer to the corresponding approximate altitude. The conversion between pressure and altitude depends on several parameters including moisture and temperature of the air. In this paper, we use the standard atmosphere, and the altitudes are therefore only an approximate altitude. To find the position of the source lightning of TGFs, we use the World Wide Lightning Location Network (WWLLN). The WWLLN matches for the TGFs are found using the same method as described in *Collier et al.* [2011]. The search found 90 matches for the first catalog TGFs and 175 matches for the second catalog. The geographic distribution of the matches from the second catalog shows that the WWLLN spherics often come from coastal areas as discussed in *Splitt et al.* [2010] and *Briggs et al.* [2013].

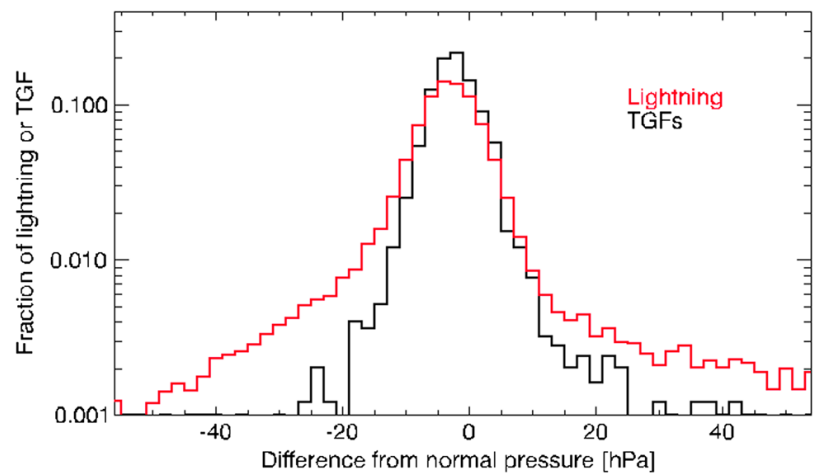

**Figure 2.** Difference between normal tropopause pressure and actual pressure for TGFs and lightning. The normal pressure is taken from the 20 year mean values of the NCEP/NCAR reanalysis. The tropopause pressure when lightning occurs is slightly lower than the normal pressure at the same place and time of year, while the pressure at the time and place of TGFs are approximately centered around the normal pressure.

For lightning data we used the individual lightning detected in 2005 and 2006 by the Lightning Imaging Sensor (LIS) on board the Tropical Rainfall Measuring Mission. Compared to WWLLN, LIS has a large and even detection efficiency within ± 35° latitude, making it suitable for statistical studies. The years 2005 and 2006 are by investigation seen to representing a typical period in terms of lightning density and geographical distribution and cover about one cycle of the quasi-biennial oscillation.

In the NCEP/NCAR data the tropopause pressure is given for every 2.5° in latitude and longitude and for every 6 h interval throughout the year. When comparing RHESSI TGFs with geolocated lightning it is found that most TGFs originate from within 400 km of the RHESSI foot point [*Hazelton et al.*, 2009; *Collier et al.*, 2011], although other TGF measurements found 300 km to be the uniform boundary [*Briggs et al.*, 2013]. For the TGFs where we have WWLLN matches, we use the point on the 2.5 × 2.5° grid that is closest to the geolocated lightning. For the other TGFs, we use the mean tropopause pressure within 400 km from the satellite foot point. An estimated error, based on a comparison between the mean and the closest point for the TGFs with geolocated lightning is found to be around 1.5 hPa.

*Smith et al.* [2010] also used the NCEP/NCAR data to find the tropopause height, but they used the monthly mean for the point closest to the foot point. The difference between the two methods is shown in Figure 1. Our method gives a wider pressure distribution, with significantly more TGFs at higher and at lower pressures.

In Figure 2, we show the difference between the tropopause pressure at the time of the TGFs and lightning compared to the normal tropopause pressure at the same place and time. The normal pressure used is the mean over 20 years for the place and time of year of the TGFs provided by the NCEP/NCAR reanalysis. The figure shows that lightning and TGFs mainly occur when the tropopause pressure is about normal. All differences larger than −20 and +10 hPa are from TGFs and lightning occurring at latitudes higher (lower) than +(−) 20° latitude.

In the tropics, the tropopause pressure is shown to have an annual pattern with relatively higher pressures in June-July-August (JJA) than in December-January-February (DJF) [*Reid and Gage*, 1996]. This variation is due to a stronger Brewer-Dobson circulation in the north subtropics winter (DJF) than in the south subtropics winter (JJA), due to more land masses in the north and the topography of these land masses [*Fueglistaler et al.*, 2009]. Since lightning and TGFs mainly happen in local summer, this means that most TGFs in the north will occur in the months when the tropopause pressure is relatively higher (JJA), and the TGFs in the south will occur when the tropopause pressure is relatively lower (DJF). The tropopause pressure for the time and place of TGFs is shown in Figure 3, which demonstrates that the tropopause pressure for TGFs in JJA is 10–15 hPa higher than in DJF between −20 and 20° latitude. If this pressure perturbation extends down to the source region for TGFs, it means that we can expect the measured TGF-to-lightning ratio to be smaller in summer (JJA) than in winter (DJF), due to an easier escape of the gamma rays from the atmosphere during December, January, and February. Unfortunately, the RHESSI statistics are not large enough for this to be investigated in the TGF data as enough TGFs in a relatively uniform geographical region is needed.

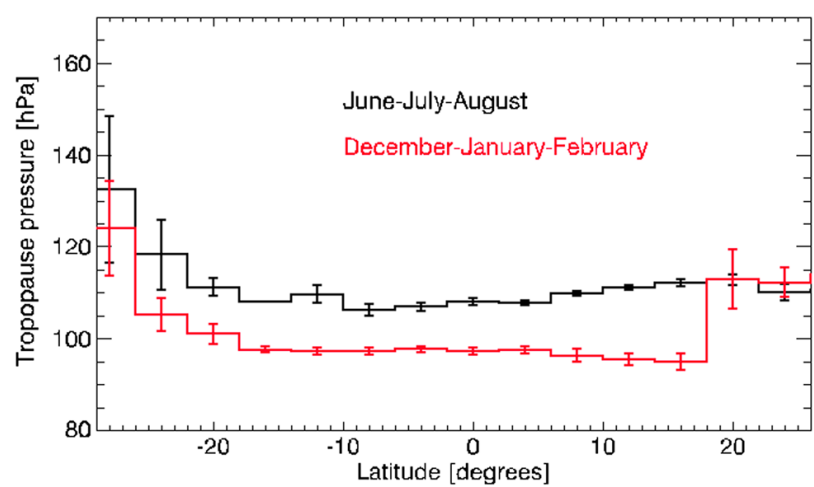

**Figure 3.** A comparison of mean tropopause pressures at the time and place of TGFs between June-July-August and in December-January-February binned by latitude. The lower pressure in DJF is due to a stronger Dobson-Brewer circulation in these months.

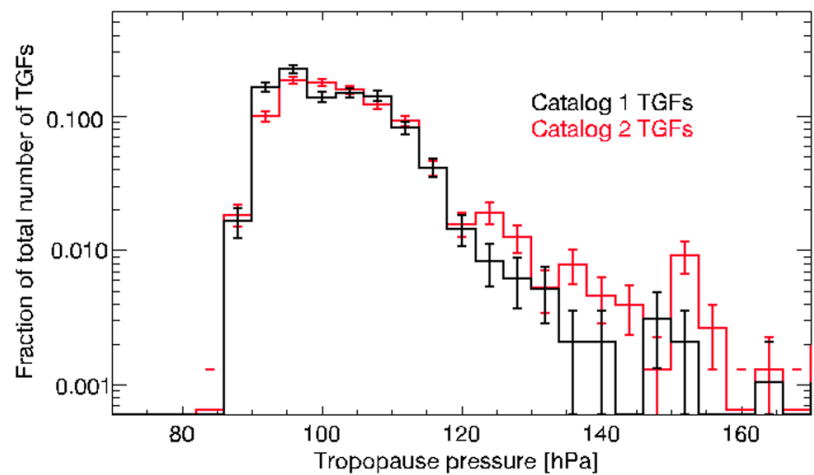

**Figure 4.** Tropopause pressure of the first catalog TGFs and the second catalog TGFs. The second catalog TGFs are mainly following the same pattern as the TGFs of the first catalog, but with a relatively larger portion of TGFs coming from times and locations where the pressure is between 125 hPa and 145 hPa.

## 3. Results

A comparison between the tropopause pressure of the TGFs in the first RHESSI catalog [*Grefenstette et al.*, 2009] and the TGFs from the second catalog [*Gjesteland et al.*, 2012] is shown in Figure 4. The figure demonstrates that the two distributions are quite similar, the main difference being between 125 hPa and 145 hPa where catalog 2 has a relatively larger number of events than catalog 1. The error bars are calculated using the square root of the number of events in each bin.

Figure 5 presents the distance between the WWLLN-located spherics and the RHESSI foot point for the 265 TGFs having a WWLLN match. The figure shows that a relatively larger number of the second catalog TGFs originates from distances between 400 and 500 km from satellite foot point than the TGFs in the first catalog. The catalog 2 TGFs that are seen to come from close distances might either be in the lower end of a power law distribution as indicated by *Østgaard et al.* [2012] or be equally strong TGFs originating from further below the tropopause.

The transmission of gamma rays in air decreases exponentially with increasing density, and the density of the atmosphere increases exponentially downward. This means that by increasing the sensitivity of RHESSI (as is done in catalog 2), it is easier to see more TGFs at larger distances than deeper in the cloud. This compares well with what we see in Figures 4 and 5.

Figure 6 shows the tropopause pressure at the time and place of lightning. We used the time and position of the individual lightnings registered by LIS in 2005 and 2006 and found the tropopause pressure of the closest point in space and time in the NCEP/NCAR data. Also plotted are the tropopause pressures for lightning of Figure 9 (black curve) of *Smith et al.* [2010]. *Smith et al.* [2010] were using a monthly map of lightning density and the monthly mean value for the tropopause pressure. It is apparent that our method gives relatively larger portion of lightning at lower tropopause pressure and slightly more spread in tropopause pressure.

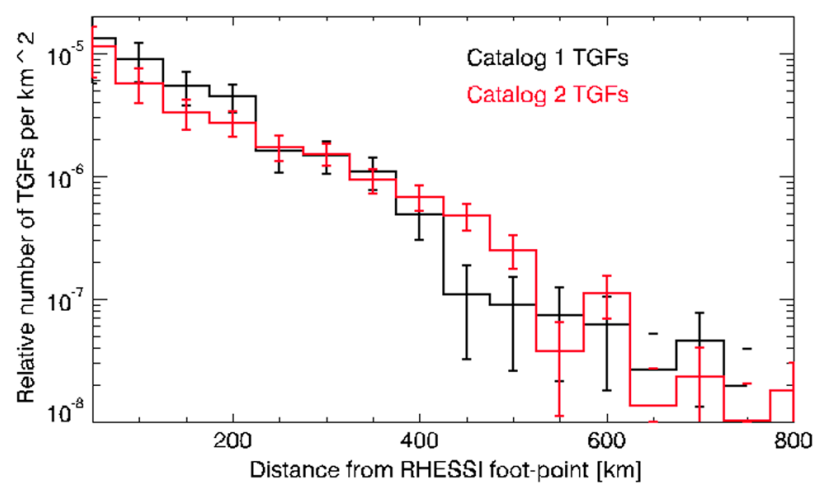

**Figure 5.** The distance between the WWLLN detected spheric and the RHESSI foot-point for the first and the second catalog TGFs. This figure shows that a relatively larger number of the catalog 2 TGFs arrive from distances between 400 and 500 km from RHESSI than the TGFs from catalog 1.

The tropopause pressures are directly related to the transmission of gamma rays from the tropopause up to satellite altitude. We used the Monte Carlo simulations described in *Østgaard et al.* [2008] to find that the number of photons escaping the atmosphere is proportional to $\exp^{\frac{-\rho_{tot}}{45[g/cm^2]}}$, where $\rho_{tot}$ is the total mass the photons are traveling through in g/cm². This is the same relation as found by *Smith et al.* [2010] and will be used to find the relative transmission coefficients for gamma rays from lightning to escape the atmosphere.

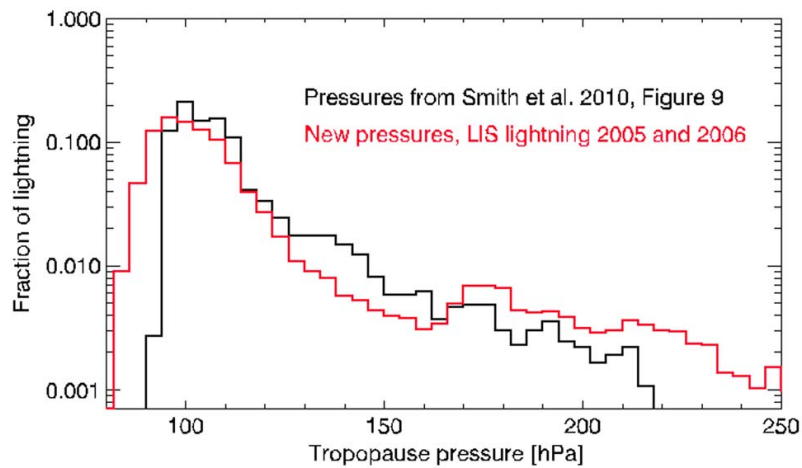

**Figure 6.** Difference in tropopause pressure at the time of lightning between the method of *Smith et al.* [2010] and our method. Our method of using 4-time daily tropopause pressures gives a generally lower pressure.

To make a comparison with the results of *Smith et al.* [2010], we made a similar geographical presentation of the difference between the transmission-corrected lightning map and the TGF map as Figure 5d of *Smith et al.* [2010]. The transmission-corrected lightning map was made by summing the transmission coefficients corresponding to each individual lightning in the LIS data from 2005 and 2006 over 2.5 × 2.5° bins divided by the total number of lightning and folded this with the LIS exposure map (sinus curve with maximum latitude of 35°) and the efficiency map of RHESSI made from the particle measurement map from *Hell and Bamberg* [2010] (Figure 2.1) and the exposure map of RHESSI (sinus curve with maximum latitude of 38°). This was then subtracted from the TGF map. As we only use relative numbers, we use arbitrary units; the values on the scale are arbitrary units.

Since our tropopause pressure is generally slightly lower than the tropopause from *Smith et al.* [2010], our probability of transmission will be higher; but because the shape of the two curves in Figure 6 is about the same, the relative difference will be fairly similar. In Figure 7, we show the difference between the first RHESSI catalog TGFs and the expected relative difference estimated from the lightning data, the exposure/efficiency and transmission. The map shows the same features as in Figure 5d of *Smith et al.* [2010], with the main discrepancies being the greater lack of TGFs in northern India and less negative values in Africa. The figure shows a similar global pattern as the inverse flash rate, duration, and radiance of LIS [*Beirle et al.*, 2014]

In Figure 8, we present the relative difference between the second RHESSI catalog TGFs and the expected relative difference between continents. This shows little or no significant difference from the first catalog TGFs, which is as expected due to the close relation in geographic distributions between the first and the second catalog of TGFs as seen in *Gjesteland et al.* [2012].

For the TGFs with WWLLN matches we know the distance from the lightning to RHESSI and the tropopause height, and we will use that to estimate the strength of the TGFs at the source using the tropopause distribution as the approximation for TGF production altitude. The source fluence can be estimated by

$$N = CN_{\text{RHESSI}}R^2 e^{\frac{\rho}{\cos\alpha 45}} \qquad (1)$$

where $N_{\text{RHESSI}}$ is the dead time-corrected number of photons measured by RHESSI, $R$ is the distance between the WWLLN spheric and RHESSI, $\rho$ is the vertical mass above the tropopause, $\alpha$ is the angle of $R$ to the vertical, and $C$ is a constant. The value of the constant $C$ is dependent on the minimum number of photons needed to get a detection in the instrument, the number of photons lost in the detector, and if the instrument is inside or outside of the initial source cone of the TGF photons. *Gjesteland et al.* [2011] Figure 3 shows a difference in the number of a factor 2 between observations made inside and outside the initial cone of the TGF photons even when accounting for the $R^2$ and $\cos(\alpha)$ effect. With an angular distribution of around 40°, we have adjusted for the angular dependence by using $C$ for $\alpha < 40$ and $2 * C$ for $\alpha > 40$. To get the numbers of $10^{17}$ and $10^{18}$ as given in literature [*Hansen et al.*, 2013; *Dwyer et al.*, 2012], the value is of the order of $10^4$.

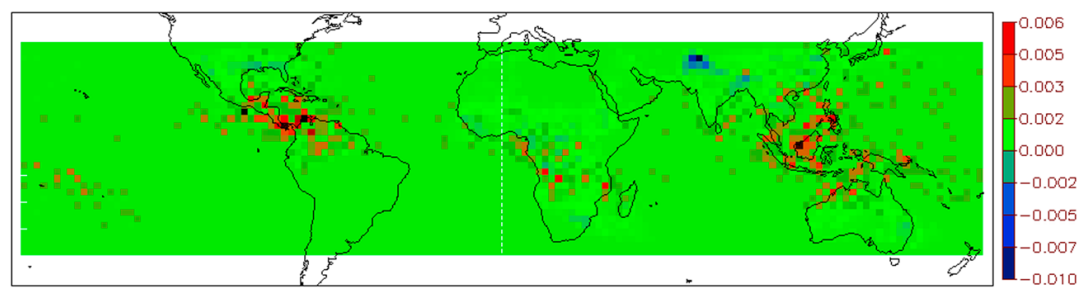

**Figure 7.** The difference between the expected number of TGFs and the measured TGFs from the first RHESSI catalog. The blue colors denote a lesser amount of TGFs than expected, and the red colors denote a bigger amount of TGFs than expected. The figure reveals a lack of TGFs in Africa compared to the expected and an excess of TGFs in Asia and America. This is the same pattern as presented in *Smith et al.* [2010].

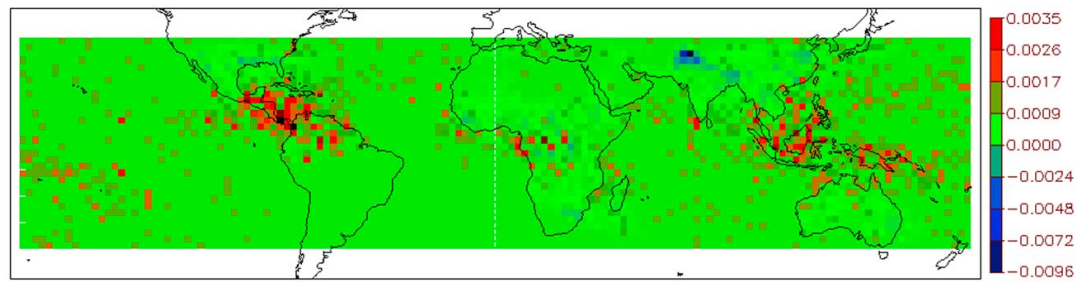

**Figure 8.** The difference between the estimated relative expected number of TGFs and the measured TGFs from the second RHESSI catalog. The blue colors denote a lesser amount of TGFs than expected, and the red colors denote a bigger amount of TGFs than expected. The second RHESSI catalog shows the same pattern as the first catalog.

The dead time-corrected fluence distribution for RHESSI of TGFs with WWLLN matches is shown in red in Figure 9. Although the distribution in Figure 9 is quite similar to what was found by *Østgaard et al.* [2012] and *Tierney et al.* [2013], a direct comparison cannot be made for the following reasons: The TGFs used in *Østgaard et al.* [2012] are only the TGFs observed by RHESSI before the degradation of the instrument in 2005–2006, while we use the TGFs measurements made up until to 2011. As the strong TGFs from after 2005 are mixed with the weaker TGFs from before 2005, this causes a softer distribution which becomes apparent when using all our TGFs. That the TGFs with WWLLN matches are comparable to the pre-2005 distribution which is due to WWLLN having higher sensitivity to strong TGFs [*Collier et al.*, 2011], making the distribution harder. The power law index found in this study cannot be directly compared to the power law index found in the previous studies, but the change in fluence distribution can be expected to be comparable to the actual change.

The black curve and line in Figure 9 are the distance and pressure-corrected fluence distribution as calculated from equation (1) and the corresponding power law. The best fit power law for the two distributions is $-2.6 \pm 0.1$ without pressure correction and $-3.2 \pm 0.1$ for the pressure-corrected TGFs. The error interval is defined so that all power law indexes with a $p$ value larger than 0.05 are within the error bars. This shows that the distribution changes slightly to a softer distribution when the altitude is accounted for.

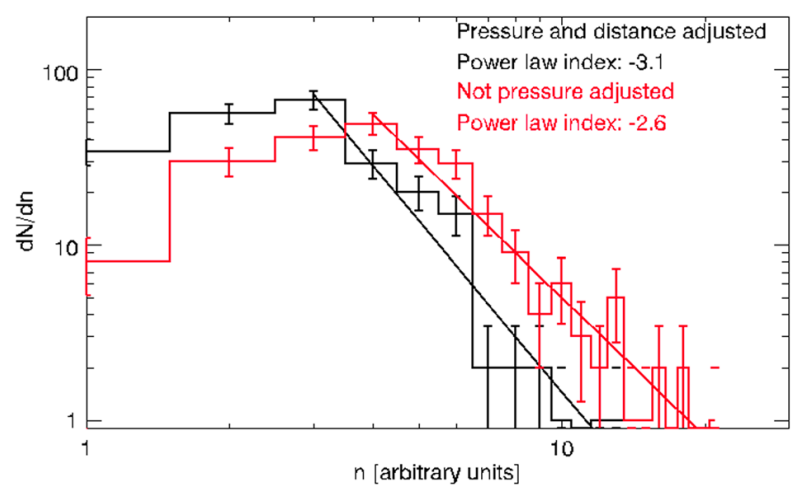

**Figure 9.** Fluence distribution of dead time-corrected RHESSI measurements with WWLLN matches and the altitude-corrected fluence distribution as calculated from equation (1), including the best fit power law for the two distributions. The pressure-corrected distribution has a slightly lower power law index than the not pressure-adjusted distribution.

*Carlson et al.* [2012] did a comparison of the source strength and the observed fluence distributions without accounting for the different altitude distributions. The results of this paper improve *Carlson et al.* [2012] analysis and demonstrate that the altitude distribution has to be considered when investigating the source fluence distribution.

## 4. Conclusion

We have shown that the TGFs of the second RHESSI catalog originate from the same altitudes as the TGFs of the first catalog, but that a bigger portion of TGFs comes from larger horizontal distances. Aside from this, the second catalog confirms the main findings of *Smith et al.* [2010]. We find that the tropopause at the time and place of TGFs is consistent with the predictions of the Brewer-Dobson circulation that the southern summer (DJF) will have a systematically lower tropopause pressure than the northern summer (JJA), and therefore, a higher fraction of TGF photons will escape the atmosphere during DJF. Based on this, we predict that the ratio of measured TGFs to lightning can be expected to have an annual variation. The main finding of this study, however, lies in the discovery that the source fluence distribution is slightly softer when the altitude distribution of TGFs is accounted for.

**Acknowledgments**
This study was supported by the European Research Council under the European Union's Seventh Framework Programme (FP7/2007-2013)/ERC grant agreement 320839 and the Research Council of Norway under contracts 208028/F50, 216872/F50, and 223252/F50 (CoE).

Alan Rodger thanks two anonymous reviewers for their assistance in evaluating this paper.